\documentclass{article}
\usepackage{amsmath}
\usepackage{amssymb}
\usepackage{latexsym}
\begin{document}

\title{An Approach to the Cosmological Constant Problem(s).
\footnote{Talk given G.L. Kane at Reoncontres de Moriond, 2004.
To appear in the proceedings of this meeting}}
\author{Gordon L. Kane\\Michigan Center for Theoretical Physics\\
University of Michigan,\\ Ann Arbor,\\ Michigan 48109,\\ USA.\\ \\
Malcolm J. Perry\\
Department of Applied Mathematics and Theoretical Physics,\\
Centre for Mathematical Sciences,\\University of Cambridge,\\
Wilberforce Road,\\
Cambridge CB3 0WA,\\ England.\\ \\
Anna N. \.Zytkow,\\ Institute of Astronomy,\\ 
University of Cambridge,\\ Madingley Road,\\ Cambridge CB3 0HA,\\
England.}

\maketitle
\vfill
\eject
\begin{abstract}
We propose an approach to explaining why  na\"\i ve large quantum
fluctuations are not the right estimate for the cosmological constant.  We
argue the universe is in a superposition of many vacua, in such a way that
the resulting fluctuations are suppressed by level repulsion to a very small
value. The approach combines several aspects of string theory and the early
history of the universe, and is only valid if several assumptions hold true.
The approach may also explain why the effective cosmological constant
remains small as the universe evolves through several phase transitions. 
It provides a non-anthropic mechanism leading to a small, non-zero
cosmological constant.
\end{abstract}

\section{Introduction}
\bigskip
There are a number of \lq\lq cosmological constant problems".

(1) The na\"\i ve quantum fluctuations in a given vacuum state are divergent and
therefore huge.  We assume that if the universe is in a supersymmetric
state the vacuum energy density vanishes. Conventionally one 
can write the vacuum energy density, which one would identify with the
density of dark energy as
\begin{equation}
\rho _{de}\sim (n_{bosons}-n_{fermions})m_{pl}^{4}+(\widetilde{m}
^{2}-m_{SM}^{2})m_{pl}^{2}+...
\end{equation}
where $\widetilde{m}$ is an appropriate supersymmetry mass scale, $m_{SM}$
is the associated Standard Model mass, and $m_{pl}$ is the Planck mass.  We
will see in the following that for our work the relevant time is at the 
beginning
of inflation, and the inflaton energy density breaks supersymmetry \cite{A}
so for us the parameter $\widetilde{m}$ will be the Hubble parameter $
H_{inf}, $ which characterizes the inflation energy density at the beginning
of inflation.  We assume $H_{inf}$ is in the range $10^{13}-$ $10^{16} GeV$ at
the beginning of inflation, a few orders of magnitude smaller than the
Planck mass. 

(2) The action for gravity plus a scalar field is

\begin{equation}
S\sim \frac{-1}{16\pi G_{N}}\int (R-2\Lambda )\sqrt{g}d^{4}x+
\int [-\frac{1}{
2}(\partial \phi )^{2}-V(\phi )]\sqrt{g}d^{4}x,
\end{equation}
so a change in the origin of the potential is equivalent to generating a
contribution to $\Lambda $ since a constant term in $V$ can just be shifted
into the first term.  This will happen at both the electroweak and QCD phase
transitions as well as any others that may exist. 
Thus one can also ask, independently of (1),
why does $\rho_{de}$ stay small ($\lesssim 3\times 10^{-3}eV)$ as the
universe evolves through several effective theories where the typical scales
are of order $100 GeV$ or $1 GeV$ or other relatively large energies?

(3) Physics also needs to explain the observed value of the cosmological
constant, $\rho _{de}^{obs}\simeq +(3\times 10^{-12}GeV)^{4},$ the infamous
\lq\lq why now" problem since $\rho _{de}^{obs}$ is similar in value to the
observed amount of baryonic matter and of dark matter.

These problems are logically separate.  Our approach 
\cite{B}, \cite{D} addresses the first two, and possibly will be relevant 
for the
third.  It is worth reminding the reader that approaches to (3) must also
explain (1) and (2) to have a full solution.  There are many approaches to
these problems, and we do not wish to criticize them.  Perhaps our approach
is complementary to some.  In particular, it could well be that our
approach could explain the first two problems, giving a very small dark
energy density, and some alternative approach could account for the observed
dark energy.

The validity of our approach will depend on the validity of several
assumptions we make.  If our approach turned out to be increasingly
convincing, it would be evidence for the validity of our assumptions. 
However, it will be clear that most or all of our assumptions can be
generalized and allow the approach to remain valid. We are focusing now on
what we hope are sufficient conditions for the approach to apply.  The
assumptions can be broadened later. \ We hope at this stage to convince the
reader that our approach is not obviously wrong, and is worthy of serious
study.

\bigskip

\section{The Basic Approach}

Here we outline the basic approach.  In the following sections we give more
details and arguments for most of the points.

$\circ $ We assume an underlying string theory, with a compactification that
includes a 6D Calabi-Yau space with N=1 supersymmetry.

$\circ $ We assume the arguments for a huge number of degenerate string
vacua \cite{C} are correct, with estimates of the number of
possible vacua being $10^{138}$ or perhaps
many more.

$\circ $ We assume that at $t=0$ three space dimensions begin to inflate,
giving an inflaton (we do not need to specify the nature of the inflaton for
this argument) energy density $\rho _{\inf }\sim H_{inf}^{2}m_{pl}^{2}$ and
breaking supersymmetry.  This will generate a potential with maxima and
minima.  We expect that the vacua will remain degenerate to within 
$T_{Hawking}\sim \sqrt{\Lambda _{\inf }}\sim H_{inf}.$  Na\"\i vely 
we expect the
potential heights and depths to be of order $H_{inf}$ since the potential
becomes flat as $H_{inf}\rightarrow 0.$  We expect $H_{inf}/m_{pl}\sim
10^{-3}$ to $10^{-6}.$  Thus na\"\i vely all the minima should be at positive
energy, though we do not have a derivation of this; our basic results do not
depend strongly on this simple picture of the potential.  This potential is
similar to but not the same as the often discussed \lq\lq landscape" 
because for
us supersymmetry is broken by the inflaton energy density and we use the
resulting potential at the beginning of inflation.

$\circ $ We expect that the wavefunction of the universe cannot be confined
to one vacuum. It must be a superposition over many vacua. 
This is the default from general principles, and any other assumption such
as expecting the universe to be in one particular vacuum would
require detailed justification.

$\circ $ We expect the vacua to mix at the beginning of inflation since the
universe is in a finite de Sitter space.  We have calculated the tunneling
rate for the de Sitter double well and shown it is not significantly
suppressed \cite{D}.  The tunneling is given in terms of the Hawking-Moss
instanton \cite{E}

$\circ $ We expect, because of the mixing, an energy spectrum to arise
analogous to the band spectrum in a solid. If $N$ vacua mix significantly
there will be $N$ levels, and the lowest will have an energy density $\sim $ 
$H_{inf}^{2}m_{pl}^{2}/N.$  One can think of this as a generalized level
repulsion -- if two levels mix, one would be lowered and the other raised,
if three mix the bottom level is lowered more, etc.

$\circ $ All this \lq\lq occurs" in the first \lq\lq instants" after 
inflation begins,
where an instant is presumably of order a few Planck times.

$\circ $ Then the universe relaxes to the ground state.  This occurs
rapidly, by several mechanisms.  It could occur before the end of
inflation, or take somewhat longer -- more thinking and analysis is needed
here.  Mechanisms include graviton pair emission, normal tunneling driven
by the Hawking temperature fluctuations, the fact that the levels have
widths, and perhaps some kind of stringy mechanisms \cite{F}.

$\circ $ It is important to understand that once the universe has relaxed to
the ground state (or near it) inflation occurs in a normal way, ending with
the release of the inflaton energy in a Big Bang. The universe is then no
longer in inflating  de Sitter space.  The finiteness of
de Sitter space is used in our mechanism only at the beginning of inflation.

$\circ $ At the electroweak and QCD phase transitions and any others, the
levels are rearranged and then the universe again relaxes to the ground
state.

$\circ $ The cosmological constant is small in the ground state, even though
it would be large perturbatively in any particular vacuum.
We emphasize that we are proposing an approach to solving the problem of why
the cosmological constant is not large from quantum fluctuations.  For our
approach to be correct our assumptions must be basically correct.  We have
made concrete assumptions as outlined above.  Most of them can probably be
generalized.  It may be that with further work the assumptions can be
independently verified -- most of them are consistent with respected
thinking in the relevant subfields. 

\bigskip

\section{What is the Residual Dark Energy? \ \lq\lq Why Now?"}

If the underlying string theory and the potential resulting from the
inflaton energy density and consequent supersymmetry breaking were
understood, we could, in principle, calculate the ground state energy density.
In practice, of course, this is not possible.  Perhaps the resulting
energy density is positive but small compared to $\rho _{de}^{obs},$ and
another source of dark energy accounts for the current accelerated expansion
of the universe, while our approach explains why the cosmological constant
is not large.  Perhaps, though, the ground state energy from our mechanism
is equal to the observed one and our mechanism can also explain "why now?".
 We cannot yet argue this, but of course we will try to find an argument
for it. 

\bigskip 

\section{Quantum Theory Analogy}

To illustrate how the energy levels arise in a calculable example, we
consider the way band spectra arise in a solid.  The calculation is
described in more detail in reference \cite{B}. 
Consider a particle of mass m moving in a periodic potential
$V(x)=V_{0}(1-\cos (2\pi x))/2.$  The bottom of each well is like a simple
harmonic oscillator, with frequency $\omega =\sqrt{2\pi ^{2}V_{0}/m}.$  If
there were only one such well it would have a ground state energy $\approx
\hbar \omega /2.$  But Bloch's theorem says the ground state wave function
is non-vanishing in all minima, not concentrated in one.  Then the mixing
generates a level repulsion and an energy spectrum. The instanton
calculation of the spectrum is well known.  Consider states $\vert n\rangle$
which are the simple harmonic oscillator states at the minimum $x=n.$  One
takes a trial Bloch wave function $\vert\theta\rangle =\sqrt{1/2\pi}
\sum_{n}e^{in\theta}\vert n\rangle $ and calculates the matrix element 
$M_{\theta ,\theta^{\prime }}=\langle \theta ^{\prime }\vert 
e^{-{{\mathcal H}t_E/\hbar}}
\vert \theta \rangle $
where ${\mathcal H}$ is the Hamiltonian and $t_E$ the Euclidean time. 
For large $t_E$,
$M_{\theta ,\theta ^{\prime }}\rightarrow \delta (\theta -\theta ^{\prime
})e^{-E(\theta ){t_E}/\hbar}$, where $E$ is the energy of the ground state. 
One finds $E(\theta )=\hbar \omega /2-2\hbar K\cos \theta e^{-S/\hbar}.$ 
$K$ is
a calculable determinant operator given by the product of the eigenvalues,
and $S=\int \sqrt{2V(x)}dx$ is the instanton action. \ The lowest energy
level occurs for $\theta =0,$ and the band has width $\Delta E=4\hbar
Ke^{-S/\hbar}.$ \ If there were $N$ discrete minima the boundaries would be the
same for $N\gg 1,$ but there would be $N$ discrete levels separated by
typically $\Delta E/N.$

\bigskip

\section{What is the Correct Generalization to the Early Universe?}

A crucial physical point for us is what is the correct generalization to the
early universe?  First, energy should be replaced by energy density for
field theory.  If we were in infinite Minkowski space the action would be
proportional to an infinite volume factor, so the factor $e^{-Vol}$ would
reduce the tunneling to zero, and no band structure would form.  But we
argue strongly that the correct physical picture to have here is that
our universe at the beginning of inflation has a positive energy
density and thus is described by de Sitter space where space is of finite
extent.  There is general agreement 
\cite{G} that for a scalar field theory with many
minima the tunneling rate is not suppressed in de Sitter spacetime, and that
the theory always has Euclidean instanton solutions such as the Hawking-Moss
one, with non-zero action.  As far as we can see the actual de Sitter
double well calculation has not previously been reported, so we have carried
out the calculation of the tunneling rate and confirmed it is indeed not
suppressed \cite{D}.  We are in the process of
extending the calculation to an asymmetric double well so we can
quantitatively discuss a potential with minima of variable depth, and also
the case with fermions so we can discuss the supersymmetric limit.

\bigskip

\section{Estimate of the Ground State Energy}

Here we follow the approach of the quantum theory example, and generalize to
the string theory type of vacuum landscape.  The calculation is described
in detail in \cite{B}  Assume the potential minima
can be described by a hypercube lattice of minima at field points
$n_{1},...n_{d},$ with each connected to $d$ others by the de Sitter tunneling,
and altogether $N$ connected by multiple transitions. If each were connected
to only a single nearest neighbor, then $d=1$. 
If all vacua could tunnel directly
to all others, then $d=N$.  With
$$
M_{\theta_{1}^{\prime },...\theta _{d}^{\prime },\theta _{1}...\theta
_{d}}=\langle \{\theta _{i}^{\prime }\}\vert 
e^{-{{\mathcal H}t_E}/\hbar}\vert
\{\theta _{i}\}\rangle \rightarrow \prod\limits_{i}\delta (\theta _{i}^{\prime
}-\theta _{i})e^{-\rho _{de}V{t_E}/\hbar} 
$$
where $V$ is the volume of space and
$$
\rho _{de}\approx H_{inf}^{2}m_{pl}^{2}-2\sum\limits_{d}H_{inf}^{4}\cos
\theta _{i}e^{-S/\hbar}. 
$$

Thus just as for the solid analogy, there is an energy spectrum.  The
spread of levels is of order $4H_{inf}^{4}d,$ and there are $N$
 levels so the
lowest level has $\rho _{de}\sim 4H_{inf}^{4}d/N.$  It is important to
understand that the tunneling here is not spacetime tunneling from one
metastable state to another, but a mechanism to calculate the energy
spectrum including non-perturbative effects, and is only used at the
beginning of inflation.

We cannot, of course, actually calculate this ground state energy at the
present time. The best we can do is to see what is required for the approach
to be consistent, which implies $\rho _{de}\leq \rho _{de}^{obs}=(3\times
10^{-12}GeV)^{4}.$  This works out if d$\leq 10^{12}$ and N$\gtrsim
10^{110}.$  In the "landscape" these are quite reasonable numbers, small
compared to the expected number of string vacua.  Next we look at the
implications of such numbers.

\bigskip

\section{The Wavefunction of the Universe is a superposition of Many Vacua!}

How can it make sense that the wave function of the universe is a
superposition of many vacua?  Can we imagine features such as the number of
families or the SM gauge group emerging as the expectation value of a
superposition?  We would like to think that the vacua that mix have certain
properties the same, perhaps in some sort of \lq\lq superselection" sense.  
These
should include the number of families, the gauge group, N=1 supersymmetry,
massless quarks and leptons, and perhaps more.  We note however that recently
it has been
argued that such quantities are not definite indices labeling string vacua.
For a recent statement of this
and references to other relevant work, see \cite{C}.
If our approach is valid such
questions are a fascinating set of issues to study.  One might imagine that
quantities such as the dilaton vev that sets the values of the gauge
couplings could more meaningfully emerge as a quantum mechanical expectation
value.  But even for this perhaps the dilaton vev is determined by a
self-dual fixed point. Our result that $N\gtrsim 10^{110}$ vacua are
superimposed could alternatively be 
stated as saying that less than $10^{-30}$ (and perhaps
far less if the larger estimates of Douglas et al are valid) of the vacua
are connected, and that seems like a reasonable result. If all the vacua
that are in the superposition share a set of properties that essentially
determine the 4D effective field theory at the string scale the goal of
calculating the detailed properties of our universe would remain a
reasonable one.

\bigskip

\section{Relaxation Time and Phase Transitions}

We expect the universe to relax to its ground state in a short time.  We
cannot yet do better than dimensional analysis for estimating the
relaxation time, but
one can list several mechanisms that seem to allow very rapid cascading. 
Firstly, each individual level will have some width which is correlated
with its time to decay to some lower energy level. Additionally,
although the results for a full calculation of the
energy spectrum would give a very complicated set of energy
levels, since  there are so many levels,
they may effectively overlap.  The presence of the Hawking temperature 
$\sim H_{inf}$ provides a thermal driving of transitions that is very large
for the higher levels.  The emission of pairs of gravitons, which will
occur on a scale of order the Planck time, should be an important mechanism.
There could also be stringy mechanisms operating as well \cite{F}.

Phenomenologically the relaxation from the initial scale of order 
$\sqrt{H_{inf}m_{pl}}$ to a scale of order $100GeV$ 
would have to occur by about $10^{-12}$ seconds 
(when the electroweak phase transition occurs), which is a
long time on the scale of Planck times.  At such a phase transition the
levels would presumably be rearranged and the universe would end up at some
energy density scale, and then continue relaxing to the ground state. 
There could be earlier transitions such as a GUT one or a
supersymmetry-breaking one, and the QCD transition occurs at a scale of
order  $1 GeV$  Our approach seems to have the capability of incorporating
these aspects of the history of the universe and explaining the apparent
fine tunings needed to understand the history. The simplest outcome would
be if the relaxation to the ground state occurred before the end of
inflation (presumably of order $10^{-35}\sec ),$ and quite rapidly after
each phase transition, but the relevant calculations have to be done to
check that.

A useful analogy may be to think of the universe as a protein folding.  The
number of minima in the landscape of the potential of a protein can be of
the same order as the number of string minima, and proteins normally find
their ground state in a millisecond or less, which compared to typical
atomic transitions suggests that it is reasonable to imagine a very rapid
transition.

\bigskip

\section{Comment on the Anthropic Principle}

The properties of our world are determined by the physics of being in the
ground state of the system of vacua, not by any particular vacuum.   For
simplicity we assume for the moment that the main global properties such as
the number of families, etc., are shared by the vacua that mix significantly
in the de Sitter space.   Then the effective 4D theory at the string scale,
which is presumably not far from the Planck scale, is (in principle)
sufficient to allow calculation of all dimensionless ratios of parameters. 
Supersymmetry breaking and mediation are determined by the Lagrangian, and
then electroweak symmetry breaking follows, all occuring with the universe
in the ground state.   We see no reason for any observable to be anthropic
-- the cosmological constant is the energy density of the ground state and
could be calculated if we knew enough about string theory, the fermion mass
ratios such as $m_{u}/m_{d}$ are determined by the superpotential, the Higgs
vev by radiative electroweak symmetry breaking (for which the $\mu $
parameter and $\tan \beta $ are calculated from the high scale theory), and
so on.   

This leads to no discomfort about why the values of some parameters (and
only some) seem to be rather near what is required for life in our universe.
  Some time ago it was realized that once inflation occurred it would be
"eternal" \cite{H} in that bubbles would inflate and
separate from any inflating universe, giving an exponential growth of the
number of universes.   Physical parameters in each of them can be different
since they depend on vacuum expectation values which can form differently,
so the many many universes will sample a range of dimensionless ratios. 
Life will occur in all universes where it can -- there is no need for
detailed probability calculations.

\bigskip

\section{Issues}

Before we could be confident our approach can explain the long-standing
problem of the apparent huge cosmological constant,
much needs to be done.   Here are some of the issues that have
to be understood better:

$\circ $ Extend the de Sitter tunneling analysis to asymmetric potentials,
to be sure the result is not unduly sensitive to fluctuations in the
potential depths and heights, and repeat the analysis with fermion fields so
the supersymmetric limit can be examined.

$\circ $ Understand the shape of the potential better, and whether the 
na\"\i ve
argument that its heights and depths are of order $H_{inf}$ is correct  for
the case of interest to us, at the beginning of inflation.

$\circ $ Understand the size of the ground state energy and whether it can
solve the \lq\lq why now" problem.

$\circ $ Understand the timescales
for the universe to relax to the ground state.

$\circ $ Understand the energy budget as a function of time: that is 
understand the energy of the ground
state,  the total energy, and relate them to energy scales later that
involve supersymmetry breaking and electroweak symmetry breaking and the QCD
condensation.

$\circ $ Understand the superposition of many minima that could provide
support for the wavefunction of the universe, and whether only vacua with
three families, etc., contribute significantly to the wavefunction.

$\circ $ Generalize the assumptions to less specific ones where possible.

$\circ $ Find phenomenological tests of specific assumptions and of the
whole picture.

If the approach is valid, these issues, while challenging, will be exciting
to study.
\bigskip

\section{Acknowledgements}

GK would like to thank the organizers for their warm hospitality,
particularly J. Tran Thanh Van and Jean-Marie Frere.   We are grateful to a
number of people for discussions and comments, A. Sevrin, L. Randall, N.
Arkani-Hamed, G. Ross, J. March-Russell, M. Einhorn, 
M. Peskin, M. Douglas, M. Duff, J. Wells, J.Liu, R. McNees and  
Q. Shafi.


\bigskip

\begin{thebibliography}{99}
\bibitem{A} This was emphasized by M.Dine, L. Randall, and S. Thomas, 
Phys. Rev. Lett. {\bf 75}(1995)398, which contains references 
to earlier work on this subject.
\bibitem{B} G. L. Kane, M. Perry, and A. \.Zytkow, hep-th/0311152. 
The basic approach
in this paper is unchanged, but we have understood some aspects and
implications better.
\bibitem{C} See M. Douglas, Statistics of string vacua, 
Proceedings of the Workshop on String Phenomenology, Durham, UK, 
hep-ph/0401004, and M. Douglas and
C.-G.Zhou, hep-th/0403018, for a recent paper and references to earlier
work. 
 The earliest estimates were by P. Candelas and X.de la Ossa, Nucl.
Phys.  {\bf B335}(1991)455.
\bibitem{D} G.L.Kane, M.Perry, and A. \.Zytkow, hep-th/0407217
\bibitem{E} I.Moss and S. Hawking, Phys. Lett. {\bf B110}(1982)35.
\bibitem{F} J.Brown and C.Teitelboim, Nucl. Phys. {\bf B297}(1988)787; 
J. Feng, J.March-Russell, S. Sethi, and F. Wilczek, 
Nucl. Phys. {\bf B602}(2001)307
\bibitem{G}See for example  M. Sasaki, E. Stewart,,
and T.Tanaka, hep-ph/9402247; S. Khlebnikov, hep-ph/0111194.
\bibitem{H} A. D. Linde, Phys. Lett. {\bf B175}(1986)395.





\end{thebibliography}
\end{document}